\documentclass[12pt]{article}             %%% Towards an explicit expression of the Seiberg-Witten map at all orders
\usepackage{amsmath,amssymb,graphicx}

\newcommand{\longlongrightarrow}{-\!\!\!-\!\!\!-\!\!\!-\!\!\!-\!\!\!\!\longrightarrow}
\newcommand{\starcom}[2]{[#1 \stackrel{\star}{,} #2]}
\newcommand{\gquad}{\qquad\qquad\qquad\qquad}

\begin{document}
\begin{titlepage}
\titlepage
\rightline{hep-th/0112027}
\rightline{CPHT-S054.1101}
\vskip 3cm
{\centering \Large \bf Towards an explicit expression\\ of the Seiberg-Witten map at all orders\\}
\vskip 1.5cm
\centerline{\bf St{\'e}phane Fidanza}
\centerline{fidanza@cpht.polytechnique.fr}
\begin{center}
\em Centre de Physique Th{\'e}orique, {\'E}cole Polytechnique\footnote{Unit{\'e} mixte du CNRS et de l'EP, UMR 7644}
\\91128 Palaiseau Cedex, France
\end{center}
\vskip 1.5cm
\begin{abstract}
The Seiberg-Witten map links noncommutative gauge theories to ordinary
gauge theories, and allows to express the noncommutative variables in
terms of the commutative ones. Its explicit form can be found order by
order in the noncommutative parameter $\theta$ and the gauge potential
$A$ by the requirement that gauge orbits are mapped on gauge
orbits. This of course leaves ambiguities, corresponding to gauge
transformations, and there is an infinity of solutions. Is there one
better, clearer than the others~? In the abelian case, we were able to
find a solution, linked by a gauge transformation to already known
formulas, which has the property of admitting a recursive formulation,
uncovering some pattern in the map. In the special case of a pure
gauge, both abelian and non-abelian, these expressions can be summed
up, and the transformation is expressed using the parametrisation in
terms of the gauge group.
\end{abstract}

\end{titlepage}

\newpage
%%%%%%%%%%%%%%%%%%%%%%%%%%%%%%%%%%%%%%%%%%%%%%%%%%%%%%%%%%%%%%%%%%%%%%%%%%%%%%%
%                                 INTRODUCTION                                %
%%%%%%%%%%%%%%%%%%%%%%%%%%%%%%%%%%%%%%%%%%%%%%%%%%%%%%%%%%%%%%%%%%%%%%%%%%%%%%%
\section{Introduction}
Noncommutativity has been studied extensively since it has become
clear that noncommutative gauge theories can describe the low energy
effective action of a brane in a constant magnetic background (see
\cite{DN,Sza} for reviews). In particular in \cite{SW}, a
correspondence, known as the Seiberg-Witten map, between gauge fields
living on D-brane worldvolume in the background of a non vanishing
constant electromagnetic field and noncommutative gauge theory on a
space with coordinate $x^\mu$ satisfying
\begin{equation}
  [x^\mu ; x^\nu ] = i \theta^{\mu\nu}
\end{equation}
has been established.\\

The simplest example of noncommutative gauge theory is the abelian
U(1), which is not a free theory, and is more similar to a U($N$)
gauge theory. In fact, at least on a torus, there exists a
transformation, the Morita equivalence, that allows to change the
noncommutativity parameter of the space and the rank of the gauge
group at the same time. For example on a 2-torus, U(1) with
noncommutativity parameter $\theta=\frac{1}{N}$ is equivalent to an
ordinary U($N$) theory. In this context, Morita equivalence can be
seen from rewriting the objects in a matrix language \cite{Sar,PS}.\\

The action of Seiberg-Witten (SW) map and Morita equivalence could be
combined, for example, to link the usual abelian Born-Infeld action to
its non-abelian version \cite{Tse}, for which the ambiguities have not yet been
solved \cite{Cor00,Cor01,Wyl,DMS}. In this process, Morita equivalence
brings all these ordering ambiguities back into the noncommutative
world, where we know that the DBI lagrangian is the lagrangian
invariant under SW map, in the slowly varying fields limit. Fixing the
ambiguities requires going beyond this limit, which means also having
a better knowledge and understanding of the SW map.\\
Another related issue is about the dynamics of the $B$ field. For
instance, if we take two snapshots of a U(1) theory, with two
different $B$ fields, they would correspond to two different
noncommutativity parameters. Let us say that the first one is linked
to an ordinary U(5) gauge group by Morita equivalence, while the
second one is related to U(3). Would that mean that the rank of the
gauge group itself has become dynamic~? We believe that there are
still things to be understood in the case of a constant $\theta$, that
could help dealing with a dynamic background.\\

In this paper, we would like to explore further the order by order
formulation of the Seiberg-Witten map, and the structures associated
with it. There already exists a formula, known as Liu's conjecture
\cite{Liu}, recently proven in \cite{LM,OO,DT,MS}, which expresses the
commutative field strength for a U(1) gauge group in terms of the
noncommutative variables non perturbatively. This formula was found by
considering the couplings of the noncommutative brane to Ramond-Ramond
potentials, and its order by order expansion involves the $\star_n$
products \cite{MW,Pal}.\\
Our approach here is completely different. It is based on solving order by
order the Seiberg-Witten equation, and is closer in spirit to
\cite{JMSSW}. Using the freedom in the possible solutions (see \cite{AK}) we
have also chosen expressions that differ from those of \cite{Pal} by a
gauge transformation. The hope was to express the map in a
simpler way, by extending the recursive formulation that appears in the usual first
orders of the field strength (see \ref{sbs:AbFS}). Our guiding criterion for
choosing a solution is to preserve its structure at all orders. Hence,
our expressions, if still order by order, can be expressed recursively
and explicitly written at any order with little work, at least in the
abelian case. Hopefully, these recursive expressions will lead to
the full non perturbative formula, as in the case of the
pure gauge, where we were able to express the solution in a particularly
suggestive form. Of course, what is interesting in the pure gauge case is
not finding a solution but rather a clue on the particular form of the
general solution.\\

The structure of the paper is as follows: definitions and notations will be found
in Section 2, as well as some remarks about our method. Section 3 contains the results
for an abelian gauge theory, and the order by order solutions are expressed
recursively. This is the main perturbative feature of the paper, and has been
explicitly checked up to order 6 (cf \ref{sbs:afd}). Work still remains
to be done to extend the pattern to all orders in $A_\mu$ (cf \ref{sbs:LineRel}).
A non perturbative explanation for the existence of this structure has been found
only in the case of the pure gauge. Then indeed, the solution can be compactly expressed
using the parametrisation in terms of the gauge group, be it abelian (cf \ref{sbs:Apg})
or not (cf \ref{sbs:NApg}). The same kind of parametrisation
may be possible in the general case, but has not yet been achieved. Some remarks
about solving the non-abelian equations are made in \ref{sbs:NAG}.\\

%%%%%%%%%%%%%%%%%%%%%%%%%%%%%%%%%%%%%%%%%%%%%%%%%%%%%%%%%%%%%%%%%%%%%%%%%%%%%%%
%                          DISCOVERING THE LANDSCAPE                          %
%%%%%%%%%%%%%%%%%%%%%%%%%%%%%%%%%%%%%%%%%%%%%%%%%%%%%%%%%%%%%%%%%%%%%%%%%%%%%%%
\setcounter{equation}{0}
\section{Discovering the landscape}
Let us consider a commutative gauge theory. The gauge group may be
general, and will not be explicit. The gauge potential is $A_\mu$, with field strength
$$ F_{\mu\nu} = \partial_\mu A_\nu - \partial_\nu A_\mu - i [A_\mu;A_\nu] $$
The gauge transformation $\delta$, with gauge parameter $\lambda$, will be acting as
$$ \delta A_\mu = \partial_\mu \lambda + i [\lambda;A_\mu] = D_\mu \lambda
   \qquad\qquad
   \delta F_{\mu\nu} = i [\lambda;F_{\mu\nu}] $$

We will first study the abelian theory, so that all commutators vanish
and the expressions simplify. When working with a non-abelian theory,
the gauge structure will then be encoded in the commutators and
anticommutators.\\

On the noncommutative side, usual multiplication of functions is
replaced by star product, which here is the Moyal-Weyl formula for
a constant noncommutativity parameter $\theta^{\mu\nu}$:
\begin{eqnarray} \label{eqn:starprod}
f \star g &=& f \; e^{\frac{i}{2}\: \overleftarrow{\partial_\rho} \theta^{\rho\sigma}
                                  \overrightarrow{\partial_\sigma}} \; g             \nonumber\\
          &=& fg + \frac{i}{2}\: (\partial_\rho f) \:\theta^{\rho\sigma}\: (\partial_\sigma g)
                 - \frac{1}{8}\: (\partial_\rho\partial_\kappa f) \:\theta^{\rho\sigma}\theta^{\kappa\tau}\:
                                 (\partial_\sigma\partial_\tau g) + O(\theta^3)
\end{eqnarray}
This allows us to write the non trivial star commutator\footnote{where we have introduced the notations
$ \{ \partial f \stackrel{\theta}{,} \partial g \} = \theta^{ij} \{ \partial_i f ; \partial_j g \} $,\\
\hspace*{1cm}
$ [ \partial\partial f \stackrel{\theta\theta}{,} \partial\partial g ]
     = \theta^{ij}\theta^{kl} [ \partial_i\partial_k f ; \partial_j\partial_l g ]$,$\quad
  \{ \partial\partial\partial f \stackrel{\theta\theta\theta}{,} \partial\partial\partial g \}
     = \theta^{ij}\theta^{kl}\theta^{mn} \{ \partial_i\partial_k\partial_m f ; \partial_j\partial_l\partial_n g \}
$.\\
The first operand carries the first index of $\theta$, making these symbols antisymmetric. Generally speaking, we will not write the indices when contractions are done in a natural way.}
\begin{equation}
i \starcom{f}{g} = i [f;g] - \frac{1}{2} \{ \partial f \stackrel{\theta}{,} \partial g \}
          - \frac{i}{8} [ \partial\partial f \stackrel{\theta\theta}{,} \partial\partial g ]
          + \frac{1}{48} \{ \partial\partial\partial f \stackrel{\theta\theta\theta}{,} \partial\partial\partial g \}
          + \dots
\end{equation}
which, in the abelian case, reduces to
\begin{equation}
i \starcom{f}{g} = - \partial f \,\theta\, \partial g
                   + \frac{1}{24}\: \partial\partial\partial f \,\theta\theta\theta\, \partial\partial\partial g
                   + \dots
\end{equation}

In particular, we have the noncommutativity of the coordinates:
$ \starcom{x^\mu}{x^\nu} = i \theta^{\mu\nu} $\\

On this space lives the noncommutative gauge theory with gauge potential $\hat{A}_\mu$ and field strength
$$ \hat{F}_{\mu\nu} = \partial_\mu \hat{A}_\nu - \partial_\nu \hat{A}_\mu - i \starcom{\hat{A}_\mu}{\hat{A}_\nu} $$
The gauge transformation $\hat{\delta}$ has gauge parameter $\hat{\lambda}$ and
$$ \hat{\delta} \hat{A}_\mu = \partial_\mu \hat{\lambda} + i \starcom{\hat{\lambda}}{\hat{A}_\mu}
                            = \hat{D}_\mu \hat{\lambda}
   \qquad\qquad
   \hat{\delta} \hat{F}_{\mu\nu} = i \starcom{\hat{\lambda}}{\hat{F}_{\mu\nu}} $$

\subsection{The Seiberg-Witten map}
The Seiberg-Witten map is a map between the commutative and the noncommutative gauge theory, which is compatible with gauge transformations. In other words, it maps gauge orbits into gauge orbits.\\
We can write noncommutative objects as functions of the commutative ones
$$ \hat{A}_\mu = \hat{A}_\mu(A_\mu;\theta) \qquad;\qquad
   \hat{F}_{\mu\nu} = \hat{F}_{\mu\nu}(A_\mu;\theta) \qquad;\qquad
   \hat{\lambda} = \hat{\lambda}(\lambda,A_\mu;\theta)
$$
and we have the following diagram, which tells that gauge transforming $\hat{A}_\mu$ as the noncommutative gauge potential or as a function of the commutative one is the same:
$$
  \begin{array}{l}
   A_\mu \stackrel{\delta}{\longlongrightarrow} A_\mu + \delta A_\mu\\
   \left\downarrow \begin{array}{c} \\ \qquad\qquad\qquad \\ \\ \end{array} \right\downarrow\\
   \hat{A}_\mu \stackrel{\hat{\delta}}{\longlongrightarrow} \hat{A}_\mu + \hat{\delta}\hat{A}_\mu\\
  \end{array}
\qquad\Leftrightarrow\qquad \widehat{A_\mu + \delta A_\mu} = \hat{A}_\mu + \hat{\delta}\hat{A}_\mu
\qquad\Leftrightarrow\qquad \delta\hat{A}_\mu = \hat{\delta}\hat{A}_\mu
$$ 
This is the Seiberg-Witten equation, which will allow us to find the map explicitly:
\begin{equation} \label{eq:SWeq}
\delta\hat{A}_\mu - \partial_\mu \hat{\lambda} = i \starcom{\hat{\lambda}}{\hat{A}_\mu}
\end{equation}

\subsection{Freedom in the solution}
To solve equation (\ref{eq:SWeq}), we develop the noncommutative variables in powers of $\theta$:
\begin{eqnarray*}
  \hat{A}_\mu &=& A_\mu + A_\mu^{(1)} + A_\mu^{(2)} + \dots\\
  \hat{F}_{\mu\nu} &=& F_{\mu\nu} + F_{\mu\nu}^{(1)} + F_{\mu\nu}^{(2)} + \dots\\
  \hat{\lambda} &=& \lambda + \lambda^{(1)} + \lambda^{(2)} + \dots
\end{eqnarray*}
Then the left hand side of equation (\ref{eq:SWeq}) is, at order $n$, exactly 
$\delta A_\mu^{(n)} - \partial_\mu \lambda^{(n)}$, while development
of the star commutator on the right hand side only involves terms of
lower (or same) order. This is the key why we can find explicit
solution to this equation, order by order in $\theta$.\\
Of course, since the only requirement was mapping gauge orbits into gauge
orbits, there is some freedom in the solution. In fact, any ``homogeneous solution",
meaning $(\mathbb{A}_\mu^{(n)};\Lambda^{(n)})$ so that
$\delta \mathbb{A}_\mu^{(n)} - \partial_\mu \Lambda^{(n)} = 0$,
can be freely added to a particular solution of the complete equation.
This ``homogeneous solution" does not change the physics, and can be rewritten
as a gauge transformation \cite{AK}. We will see an example of it in section \ref{sbs:ab2o},
where we relate our solution with the already known solutions of \cite{Liu,OO}.\\

\begin{figure}[htbp]
\centering
\includegraphics{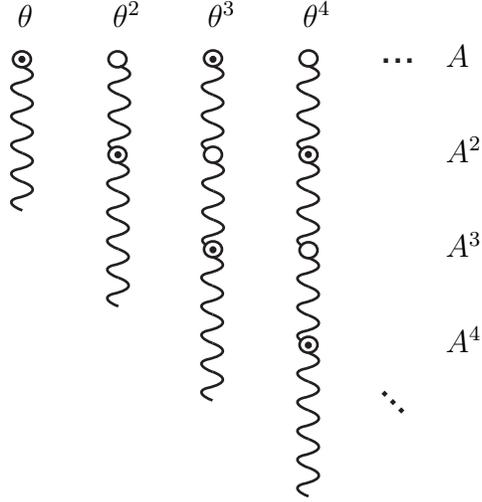}
\caption{\small Circles are non-abelian terms, that spread on higher orders in $A$, hence the tails. When the group is abelian, many terms drop out, leaving only the dots, which do not spread anymore.}
\label{fig:Diag}
\begin{picture}(0,0)(0,0)
\put(85,215){$A$}
\put(85,179){$A^2$}
\put(85,143){$A^3$}
\put(85,107){$A^4$}
\put(-77,230){$\theta$}
\put(-41,230){$\theta^2$}
\put(-05,230){$\theta^3$}
\put( 31,230){$\theta^4$}
\end{picture}
\end{figure}

In the following, there will be another type of development: order by
order in powers of $A$ (fig.\ref{fig:Diag}). In fact, each term in a
solution can only involve, algebraically, $A$, $\theta$ and
derivatives. Specifying the number of $A$ and the number of $\theta$
then characterizes a class of terms (freedom remains in the
contraction of indices, and the action of derivatives). We will denote
a term of order $(A^n;\theta^m)$ in the development of $\hat{\lambda}$
or $\hat{A}_\mu$ by $\lambda^{(n,m)}$ or $A_\mu^{(n,m)}$.\\
Let us stress that this order in $A$, whereas well defined in the
abelian theory, becomes rather unclear in a non-abelian theory. The
field strength, the covariant derivative and the gauge transformation
contain a quadratic part in $A$. This does not mean that order in $A$
does not make sense, but that we will have to be very careful in the
identification of the terms. Misidentification would spoil the
picture.\\
We will also see that it may be more efficient to sum all the terms of
a given order in $A$, and not in $\theta$, as advertised. Thus, we
will use the notation $\lambda^{(n,\infty)}$, $A_\mu^{(n,\infty)}$
meaning order $n$ in $A$
$$
 \lambda^{(n,\infty)} = \sum_{m=0}^{\infty} \lambda^{(n,m)} \qquad;\qquad
 A_\mu^{(n,\infty)}   = \sum_{m=0}^{\infty} A_\mu^{(n,m)}
$$

%%%%%%%%%%%%%%%%%%%%%%%%%%%%%%%%%%%%%%%%%%%%%%%%%%%%%%%%%%%%%%%%%%%%%%%%%%%%%%%
%                                ABELIAN THEORY                               %
%%%%%%%%%%%%%%%%%%%%%%%%%%%%%%%%%%%%%%%%%%%%%%%%%%%%%%%%%%%%%%%%%%%%%%%%%%%%%%%
\setcounter{equation}{0}
\section{The abelian theory}
In the abelian theory, different orders in $A$ do not mix together. We will start with the first diagonal of the diagram: the terms of order $(A^n,\theta^n)$. We will see that we can write solutions so that each point is given from the previous one. The expressions can then also be written easily without recursivity. The emphasized expressions have been checked up to and including $n=6$. For $n=4,5,6$, this has been done using a computer program specially written on this purpose.

\subsection{The first diagonal: orders $(A^n,\theta^n)$} \label{sbs:afd}
\subsubsection{First order}
The first order equation is reduced to
$\quad \delta A_\mu^{(1,1)} - \partial_\mu \lambda^{(1,1)} = -\partial\lambda \theta \partial A_\mu \quad$
and admits the usual solution \cite{SW}
\begin{eqnarray*}
  \lambda^{(1,1)} &=& -\frac{1}{2}\: \theta^{\rho\sigma} A_\rho \partial_\sigma \lambda \\
  A_\mu^{(1,1)}   &=& -\frac{1}{2}\: \theta^{\rho\sigma} A_\rho (\partial_\sigma A_\mu + F_{\sigma\mu})
\end{eqnarray*}
We can rewrite this solution to make its structure more apparent, so that we can recognize the form of next orders
\begin{eqnarray} \label{eqn:ord1}
  \lambda^{(1,1)} &=& -\frac{1}{2}\: (A\theta \partial) \lambda \nonumber\\
  A_\mu^{(1,1)}   &=& -\frac{1}{2}\: (A\theta \partial) A_\mu -\frac{1}{2}\: (A\theta F)_\mu
\end{eqnarray}

\subsubsection{Second order} \label{sbs:ab2o}
The equation at order $(A^2,\theta^2)$ is
$\displaystyle{
   \delta A_\mu^{(2,2)} - \partial_\mu \lambda^{(2,2)} = - \partial\lambda^{(1,1)} \theta \partial A_\mu
                                                         - \partial\lambda \theta \partial A_\mu^{(1,1)} }$.\\
Solving with the previous (1,1) expressions, we get the solution
\begin{eqnarray} \label{eqn:ord2}
 \lambda^{(2,2)} &=& \frac{1}{6}\: [ \: (A\theta\partial)(A\theta\partial) \lambda
                                       + A\theta F \theta \partial \lambda           \: ] \nonumber\\
 A_\mu^{(2,2)}   &=& \frac{1}{6}\: [ \: (A\theta\partial)(A\theta\partial) A_\mu
                                       + A\theta F \theta \partial A_\mu
                                       + 2 (A\theta\partial)(A\theta F)_\mu
                                       + 2  A\theta F \theta F_\mu                   \: ]
\end{eqnarray}
But there are other possibilities. In \cite{JMSSW} another solution was found
\begin{eqnarray*}
 {\lambda^{(2,2)}}_{\rm JMSSW} &=& \frac{1}{2}\: \theta^{\rho\sigma} \theta^{\kappa\tau} A_\rho
        \partial_\sigma A_\kappa \partial_\tau \lambda \\
 {A_\mu^{(2,2)}}_{\rm JMSSW}   &=& \frac{1}{2}\: \theta^{\rho\sigma} \theta^{\kappa\tau} A_\rho
        [ \partial_\sigma A_\kappa \partial_\tau A_\mu + A_\kappa \partial_\sigma F_{\tau\mu}
        + F_{\sigma\kappa} F_{\tau\mu} ]
\end{eqnarray*}
The difference between the two is the ``homogeneous solution" $\frac{1}{6}H$, where $H$ is
$$ \begin{array}{rclcr}
 \Lambda^{(2,2)}        &=& \theta^{\rho\sigma} \theta^{\kappa\tau} A_\rho
        [ \partial_\kappa A_\sigma + \partial_\sigma A_\kappa - A_\kappa \partial_\sigma ] \partial_\tau \lambda
                        &=& \delta [A\theta(\partial A)\theta A]\\[2mm]
 \mathbb{A}_\mu^{(2,2)} &=& \theta^{\rho\sigma} \theta^{\kappa\tau} A_\rho
        [ \partial_\kappa A_\sigma + \partial_\sigma A_\kappa - A_\kappa \partial_\sigma ] \partial_\mu A_\tau
                        &=& \partial_\mu [A\theta(\partial A)\theta A]
\end{array} $$
In particular, it is a gauge transformation on $\hat{A}_\mu$, with parameter $\frac{1}{6} A\theta(\partial A)\theta A$.\\
The solution (\ref{eqn:ord2}) is also different from the order 2 expansion of Liu's formula \cite{Liu,MW,Pal}. This time, the difference is $\frac{5}{12}H$, and it is again a gauge transformation. Of course, in both cases, the field strength is the same.

\subsubsection{Third order}
The third order begins to get very messy. The equation is
$$ \delta A_\mu^{(3,3)} - \partial_\mu \lambda^{(3,3)} = - \partial\lambda^{(2,2)} \theta \partial A_\mu
 - \partial\lambda^{(1,1)} \theta \partial A_\mu^{(1,1)} - \partial\lambda \theta \partial A_\mu^{(2,2)} $$
Expanding it with the actual expressions (\ref{eqn:ord1}) and (\ref{eqn:ord2}) gives a great number of terms. But we can build an expression on the model arising from (\ref{eqn:ord1}) and (\ref{eqn:ord2}). And indeed, we explicitly checked that the following is a solution to this equation
\begin{eqnarray} \label{eqn:ord3}
 \lambda^{(3,3)} &=& -\frac{1}{24}\: [ \:
   (A\theta\partial)(A\theta\partial)(A\theta\partial) \lambda + (A\theta F \theta\partial)(A\theta\partial) \lambda
                                       \nonumber\\
  &&\gquad     + 2 (A\theta\partial)(A\theta F \theta\partial) \lambda + 2 A\theta F \theta F \theta\partial \lambda
                                  \: ] \nonumber\\
 A_\mu^{(3,3)}   &=& -\frac{1}{24}\: [ \:
     (A\theta\partial)(A\theta\partial)(A\theta\partial) A_\mu + (A\theta F \theta\partial)(A\theta\partial) A_\mu
                                       \nonumber\\
  &&\gquad       + 2 (A\theta\partial)(A\theta F \theta\partial) A_\mu + 2 A\theta F \theta F \theta\partial A_\mu
                                       \nonumber  \\
  &&\qquad     + 3 (A\theta\partial)(A\theta\partial)(A\theta F)_\mu + 3 (A\theta F \theta\partial)(A\theta F)_\mu
                                       \nonumber \\
  &&\gquad                       + 6 (A\theta\partial)(A\theta F \theta F)_\mu + 6 A\theta F \theta F \theta F_\mu
                                  \: ]
\end{eqnarray}
The interest of having this explicit solution is that we see a
structure emerging. The operators look very similar, the coefficients
seem to follow some rule. The first two orders were not really enough to
draw conclusions, but order 3 strongly suggests that the form can be
generalized to a solution of order $n$. Furthermore, we developed a computer
code which allowed us to check that these expressions are also solutions
at order 4, 5 and 6.\\

\subsubsection{Recursive formulation}
Better than describing some prescriptions leading to the general form
of order $n$ -- or rather $(n,n)$, we can express the results obtained
so far in a recursive manner.\\
Recalling the first order solution
\begin{eqnarray*}
 \lambda^{(1,1)} &=& -\frac{1}{2}\: A\theta \partial\lambda \\
 A_\mu^{(1,1)}   &=& -\frac{1}{2}\: A\theta (\partial A_\mu + F_{\cdot\mu})
\end{eqnarray*}
and comparing to (\ref{eqn:ord2}) and (\ref{eqn:ord3}), we can write the second and third orders
in a similar way, with $A^{(1,1)}$ or $A^{(2,2)}$ {\em acting as
  operators}, under the form given above. That means explicit
derivatives (contrary to those hidden in $F$s) also act on what is on
the right.
$$ \begin{array}{rcl}
%\begin{eqnarray*}
 \lambda^{(2,2)} &=& -\frac{1}{3}\: A^{(1,1)}\theta \partial\lambda \\[2mm]
 A_\mu^{(2,2)}   &=& -\frac{1}{3}\: A^{(1,1)}\theta (\partial A_\mu + 2 F_{\cdot\mu})
%\end{eqnarray*}
   \end{array}
\qquad
   \begin{array}{rcl}
%\begin{eqnarray*}
 \lambda^{(3,3)} &=& -\frac{1}{4}\: A^{(2,2)}\theta \partial\lambda \\[2mm]
 A_\mu^{(3,3)}   &=& -\frac{1}{4}\: A^{(2,2)}\theta (\partial A_\mu + 3 F_{\cdot\mu})
%\end{eqnarray*}
   \end{array}
$$
This leads to the conjecture that the same iterative relation gives a solution for general $n$
\begin{eqnarray} \label{eq:AbRec}
 \lambda^{(n,n)} &=& -\frac{1}{n+1}\: A^{(n-1,n-1)}\theta \partial\lambda \nonumber\\
 A_\mu^{(n,n)}   &=& -\frac{1}{n+1}\: A^{(n-1,n-1)}\theta (\partial A_\mu + n F_{\cdot\mu})
\end{eqnarray}

It is also possible to express an algebraically factorized form
without recursivity. Let us denote as $d$ and $f$ the two basic blocks
that are building all the terms
$$ d \equiv A\theta\partial    \qquad\qquad    f \equiv A\theta F A^{-1} $$
with $A_\rho^{-1} A_\sigma = \delta_{\rho\sigma}$ and the rule
$A^{-1}\lambda=0$ (or $f\lambda=0$). Then
\begin{eqnarray*}
 \lambda^{(n,n)} &=& \frac{(-1)^n}{(n+1)!}\; (d+f) (d+2f) \dots (d+nf) \; \lambda \\
 A_\mu^{(n,n)}   &=& \frac{(-1)^n}{(n+1)!}\; (d+f) (d+2f) \dots (d+nf) \; A_\mu
\end{eqnarray*}
The only effect of the $f\lambda=0$ rule is that when developing the last factor, the formulas for $A$ have twice the number  of terms than those for $\lambda$, as has been seen previously.\\
It might be interesting to learn more about the significance and
properties of these $f$ and $d$, even if they can be thought of as the
first order of more general objects. Indeed, we will see the first
correction in section \ref{sbs:LineRel}.

\subsubsection{The field strength} \label{sbs:AbFS}
We calculated the noncommutative field strength corresponding to the previous expressions for $\hat{A_\mu}$
\begin{eqnarray*}
 F_{\mu\nu}^{(1,1)} &=& \partial_\mu A_\nu^{(1,1)} - \partial_\nu A_\mu^{(1,1)} + \partial A_\mu \theta \partial A_\nu
                   \,=\, -(A\theta\partial)F_{\mu\nu} - F_{\mu\cdot} \theta F_{\cdot\nu} \\
 F_{\mu\nu}^{(2,2)} &=& \partial_\mu A_\nu^{(2,2)} - \partial_\nu A_\mu^{(2,2)}
                      + \partial A_\mu^{(1,1)} \theta \partial A_\nu + \partial A_\mu \theta \partial A_\nu^{(1,1)}\\
                    &=& \frac{1}{2}(A\theta\partial)(A\theta\partial)F_{\mu\nu}
                      + \frac{1}{2}(A\theta F \theta\partial)F_{\mu\nu}
                      + (A\theta\partial)(F_{\mu\cdot} \theta F_{\cdot\nu})
                      + F_{\mu\cdot} \theta F \theta F_{\cdot\nu} \\
 F_{\mu\nu}^{(3,3)} &=& \partial_\mu A_\nu^{(3,3)} - \partial_\nu A_\mu^{(3,3)}
                      + \partial A_\mu^{(2,2)} \theta \partial A_\nu + \partial A_\mu \theta \partial A_\nu^{(2,2)}
                      + \partial A_\mu^{(1,1)} \theta \partial A_\nu^{(1,1)} \\
                    &=& -\frac{1}{6} \left[ \begin{array}{ccc}
           (A\theta\partial)(A\theta\partial)(A\theta\partial)F_{\mu\nu}
      &+& 2(A\theta\partial)(A\theta F\theta\partial)F_{\mu\nu}\\
          +(A\theta F\theta\partial)(A\theta\partial)F_{\mu\nu} &+& 2(A\theta F\theta F\theta\partial)F_{\mu\nu}
                                     \end{array} \right]\\
                    & & -\frac{1}{2} \left[ \begin{array}{l}
           (A\theta\partial)(A\theta\partial)(F_{\mu\cdot} \theta F_{\cdot\nu})\\
          +(A\theta F\theta\partial)(F_{\mu\cdot} \theta F_{\cdot\nu})
                                     \end{array} \right] - \left[ \begin{array}{l}
           (A\theta\partial)(F_{\mu\cdot} \theta F \theta F_{\cdot\nu})\\
          + F_{\mu\cdot} \theta F \theta F \theta F_{\cdot\nu}
                                     \end{array} \right]
\end{eqnarray*}
And again we identify in these expressions the same recursive pattern
\begin{eqnarray} \label{eq:AbFRec}
 F_{\mu\nu}^{(1,1)} &=& -(A\theta\partial)F_{\mu\nu} - F_{\mu\cdot} \theta F_{\cdot\nu} \nonumber\\
 F_{\mu\nu}^{(2,2)} &=& -(A^{(1,1)}\theta\partial)F_{\mu\nu} - F_{\mu\cdot}^{(1,1)} \theta F_{\cdot\nu} \nonumber\\
 F_{\mu\nu}^{(3,3)} &=& -(A^{(2,2)}\theta\partial)F_{\mu\nu} - F_{\mu\cdot}^{(2,2)} \theta F_{\cdot\nu}
\end{eqnarray}
Let us stress that the second line only involves usual formulas:
$F_{\mu\nu}^{(1,1)}$, $F_{\mu\nu}^{(2,2)}$ and $A_\mu^{(1,1)}$ are the same
here as in \cite{Liu,JMSSW}. The particular choice that we have made for
$A_\mu^{(2,2)}$ and others then allowed the same pattern for the third line.

\subsection{A line by line relation~?} \label{sbs:LineRel}
Now that we have some expression for the first diagonal, it still
remains to fill the entire triangle on fig.\ref{fig:Diag} to get all
the terms. In fact, the whole first line (order 1 in $A$) is really
easy to find, and the solution is essentially unique. It is known as a
$\star_2$ or $\star'$ product expression \cite{MW}:
$$ A_\mu^{(1,\infty)} = -\frac{1}{2} \theta^{ij} A_i \star_2 (\partial_j A_\mu + F_{j\mu})
     \qquad\mbox{with}\quad
   f(x) \star_2 g(x) =
    \frac{\sin(\frac{\partial_1 \theta \partial_2}{2})}{\frac{\partial_1 \theta \partial_2}{2}} f(x_1) g(x_2) |_{x_i=x}
$$
which have expansion in $\theta$
\begin{eqnarray*}
A_\mu^{(1,\infty)} &=& A_\mu^{(1,1)} + A_\mu^{(1,3)} + \dots\\
A_\mu^{(1,\infty)} &=& \frac{1}{2} \left[ -A \theta (\partial A_\mu + F_{\cdot\mu})
          + \frac{1}{24}\:\partial\partial A \:\theta\theta\theta\: \partial\partial (\partial A_\mu + F_{\cdot\mu})
          + \dots \right]
\end{eqnarray*}

We want to use this to extend the previous
recursive relation, holding for the first diagonal, to a line by line
relation. The basic reason to try to sum by lines (see
fig.\ref{fig:Diag}), is that all terms of the first line depend on
$\partial A_\mu + F_{\cdot\mu}$, while terms of order $(n,n)$ are
expected to depend on $\partial A_\mu + n F_{\cdot\mu}$, as can be
seen on (\ref{eq:AbRec}). We have calculated a solution at order
(2,4), the second term on the second line, and indeed it happens to
depend on $\partial A_\mu + 2 F_{\cdot\mu}$.\\
Then, one may wonder about the extended recursive rules. Recall that
the expression at order 2 was the same as the solution for order 1,
with
\begin{itemize}
\item the starting $A$ replaced by $A^{(1,1)}$
\item $\displaystyle{\frac{\partial A_\mu + F_{\cdot\mu}}{2}}$
          changed to $\displaystyle{\frac{\partial A_\mu + 2 F_{\cdot\mu}}{3}}$
\end{itemize}
If we do the same for all the line, we get
\begin{eqnarray*}
\lambda^{(1,\infty)} &=& \frac{1}{2} \left[ -A \theta \partial\lambda
          + \frac{1}{24} \partial\partial A \:\theta\theta\theta\: \partial\partial\partial\lambda + \dots \right]\\
\lambda^{(2,\infty)} &=& \frac{1}{3} \left[ -A^{(1,\infty)} \theta \partial\lambda
          + \frac{1}{24}\:\partial\partial A^{(1,\infty)} \:\theta\theta\theta\: \partial\partial\partial\lambda
          + \dots \right]\\
    &=& \underbrace{-\frac{1}{3} A^{(1,1)} \theta \partial\lambda}_{\lambda^{(2,2)}} \:
        \underbrace{-\frac{1}{3} A^{(1,3)} \theta \partial\lambda
                    +\frac{1}{72} \partial\partial A^{(1,1)} \:\theta\theta\theta\: \partial\partial\partial\lambda
                   }_{\lambda^{(2,4)}}
        + \dots
\end{eqnarray*}
And the same for $A_\mu^{(2,\infty)}$, replacing $\partial\lambda$ by
$(\partial A_\mu + 2 F_{\cdot\mu})$. Do not forget that the
$A^{(1,m)}$ have to act as operators (the rest of the term is ``glued"
inside the action of their derivatives). This prescription leads to
ambiguities\footnote{
We shall try to comment on this here. In the term
$$ \partial\partial A^{(1,1)} \:\theta\theta\theta\: \partial\partial\partial\lambda \propto
   (A \theta \overrightarrow{\partial}) A \:
   (\overleftarrow{\partial\partial} \:\theta\theta\theta\: \overrightarrow{\partial\partial\partial}\lambda) $$
$A^{(1,1)}$ should act as an operator on the right, while
$\overleftarrow{\partial\partial}$ is acting on the left. Note that
this is the first instance where such a situation occurs. One will
keep meeting similar ambiguities at higher orders.
}, up to which this proposal for order (2,4) reproduces an actual solution
(out of a dozen, all but one terms are reproduced unambiguously).\\

The same should also hold for the field strength. Generalizing the
formulas (\ref{eq:AbFRec}) would give $F_{\mu\nu}^{(2,\infty)}$ from
$F_{\mu\nu}^{(1,\infty)}$, with
$$
F_{\mu\nu}^{(1,\infty)} = \left[ \begin{array}{ccccc}
   -A \theta \partial F_{\mu\nu} &+& \displaystyle
         \frac{1}{24}\:\partial\partial A \:\theta\theta\theta\: \partial\partial\partial F_{\mu\nu} &+& \dots\\[3mm]
   - F_{\mu\cdot} \theta F_{\cdot\nu} &+& \displaystyle
         \frac{1}{24}\:\partial\partial F_{\mu\cdot} \:\theta\theta\theta\: \partial\partial F_{\cdot\nu} &+& \dots
                          \end{array} \right]
$$

\subsection{Pure gauge case} \label{sbs:Apg}
We consider here the case when the gauge potential $A$ is a pure
gauge. For an abelian theory, this can be done with $A_\rho =
\partial_\rho \alpha$. As expected, the field strength vanishes, and
the gauge transformation of $\alpha$ should be $\delta \alpha =
\lambda$.\\

The objective here is not to find the solutions: it is the orbit of
the noncommutative pure gauge. Indeed, we expect from the map that
zero is mapped to zero, and this statement can only hold for the whole
orbits.\\
What we are looking for is an expression that, in some sense, is nicer
than the others. The solutions emphasized up to now will indeed greatly
simplify in the pure gauge case. That will allow two different improvements,
which are still out of reach for a general potential:
first, the solutions have a natural non-abelian generalisation, and second, they
will sum up to a non-perturbative form (as we will see in \ref{sbs:NApg}).
This suggests that the same kind of expressions could hold for general
potentials, once they are parametrized by elements of the gauge group.\\

\subsubsection{The solution for $A$ and $\lambda$} \label{sbs:abpg}
We can rewrite the solution at order $(1,\infty)$
$$ A_\mu^{(1,\infty)} = -\frac{1}{2} \partial\alpha \theta \partial A_\mu
          + \frac{1}{48} \partial\partial\partial\alpha \:\theta\theta\theta\: \partial\partial\partial A_\mu + \dots
$$

and recognize the development of a shorter formula (the same holds for $\lambda^{(1,\infty)}$)
\begin{eqnarray*}
  A_\mu^{(1,\infty)}   &=& \frac{i}{2} \starcom{\alpha}{A_\mu}\\
  \lambda^{(1,\infty)} &=& \frac{i}{2} \starcom{\alpha}{\lambda}
\end{eqnarray*}

The Seiberg-Witten equation, written at each order in $A$, gives\footnote{
From now on, we drop the $\infty$ since no confusion can arise: $A_\mu^{(n)}$ means $A_\mu^{(n,\infty)}$}
$$
\begin{array}{llrcl}
 \mbox{Order }A &:\qquad& \delta A_\mu^{(1)} - \partial_\mu \lambda^{(1)}
    &=& i \starcom{\lambda}{A_\mu}\\
 \mbox{Order }A^2 &:& \delta A_\mu^{(2)} - \partial_\mu \lambda^{(2)}
    &=& i \starcom{\lambda}{A_\mu^{(1)}} + i \starcom{\lambda^{(1)}}{A_\mu}\\
 \mbox{Order }A^n &:& \delta A_\mu^{(n)} - \partial_\mu \lambda^{(n)}
    &=& \displaystyle \sum_{p=0}^{n-1} i \starcom{\lambda^{(p)}}{A_\mu^{(n-1-p)}}
\end{array}
$$

Now we can check directly that the expression is indeed solution at order $A$
\begin{eqnarray*}
  \delta A_\mu^{(1)} - \partial_\mu \lambda^{(1)}
    &=& \frac{1}{2}i \starcom{\lambda}{A_\mu} + \frac{1}{2}i \starcom{\alpha}{\partial_\mu \lambda}
      - \frac{1}{2}i \starcom{\partial_\mu \alpha}{\lambda} - \frac{1}{2}i \starcom{\alpha}{\partial_\mu \lambda} \\
    &=& i \starcom{\lambda}{A_\mu}
    \quad \equiv \quad i \starcom{\lambda^{(0)}}{A_\mu^{(0)}}
\end{eqnarray*}

This formula is still valid to all orders, as we shall now prove
recursively. Suppose we have the solution for $k \leq n$
\begin{eqnarray*}
  A_\mu^{(k)}   &=& \frac{i}{k+1} \starcom{\alpha}{A_\mu^{(k-1)}}\\
  \lambda^{(k)} &=& \frac{i}{k+1} \starcom{\alpha}{\lambda^{(k-1)}}
\end{eqnarray*}
and define order $(n+1)$ by the same recursive formula. Then
$$
  \delta A_\mu^{(n+1)} - \partial_\mu \lambda^{(n+1)}
    = \frac{i}{n+2} \starcom{\lambda}{A_\mu^{(n)}} + \frac{i}{n+2} \starcom{\lambda^{(n)}}{A_\mu}
    + \frac{i}{n+2} \starcom{\alpha}{\delta A_\mu^{(n)} - \partial_\mu \lambda^{(n)}}
$$
and with the help of the Jacobi identity
\begin{eqnarray*}
  \starcom{\alpha}{\delta A_\mu^{(n)} - \partial_\mu \lambda^{(n)}}
    &=& \starcom{\alpha}{\sum_{p=0}^{n-1} i \starcom{\lambda^{(p)}}{A_\mu^{(n-1-p)}}}\\
    &=& \sum_{p=0}^{n-1} \left( \starcom{\lambda^{(p)}}{i \starcom{\alpha}{A_\mu^{(n-1-p)}}}
                              - \starcom{A_\mu^{(n-1-p)}}{i \starcom{\alpha}{\lambda^{(p)}}} \right)\\
    &=& \sum_{p=0}^{n-1} \left( (n+1-p) \starcom{\lambda^{(p)}}{A_\mu^{(n-p)}}
                              - (p+2) \starcom{A_\mu^{(n-1-p)}}{\lambda^{(p+1)}} \right)
\end{eqnarray*}
which is exactly what is needed to finally give
$$ \delta A_\mu^{(n+1)} - \partial_\mu \lambda^{(n+1)} = \sum_{p=0}^{n} i \starcom{\lambda^{(p)}}{A_\mu^{(n-p)}} $$

So we arrive at a solution to all orders for $\hat{A}_\mu$ and
$\hat{\lambda}$, in the form
\begin{equation} \label{eq:AbRecPG}
  \begin{array}{rcl} \vspace{2mm}
    A_\mu^{(n+1)}   &=& \displaystyle{\frac{i}{n+2}\: \starcom{\alpha}{A_\mu^{(n)}}}\\
    \lambda^{(n+1)} &=& \displaystyle{\frac{i}{n+2}\: \starcom{\alpha}{\lambda^{(n)}}}
  \end{array}
\end{equation}
$A_\mu^{(0)}$ and $\lambda^{(0)}$ being $A_\mu$ and $\lambda$.

\subsubsection{The field strength}
Here we compute the field strength corresponding to the solution
(\ref{eq:AbRecPG}) for the noncommutative connection, and show
recursively that it indeed vanishes. $\hat{F}_{\mu\nu}$ can be
developed in powers of $A$, giving
\begin{eqnarray*}
  F_{\mu\nu}^{(0)} &\equiv& F_{\mu\nu} = 0\\
  F_{\mu\nu}^{(1)} &\equiv& \partial_\mu A_\nu^{(1)} - \partial_\nu A_\mu^{(1)} - i \starcom{A_\mu}{A_\nu}\\
                   &=& \frac{i}{2}\starcom{A_\mu}{A_\nu} + \frac{i}{2}\starcom{\alpha}{\partial_\mu A_\nu}
                     - \frac{i}{2}\starcom{A_\nu}{A_\mu} - \frac{i}{2}\starcom{\alpha}{\partial_\nu A_\mu}
                     - i \starcom{A_\mu}{A_\nu}\\
                   &=& \frac{i}{2}\starcom{\alpha}{F_{\mu\nu}} = 0\\
  F_{\mu\nu}^{(n+1)} &\equiv& \partial_\mu A_\nu^{(n+1)} - \partial_\nu A_\mu^{(n+1)}
                            - \sum_{k=0}^n i \starcom{A_\mu^{(k)}}{A_\nu^{(n-k)}}\\
                     &=& \frac{i}{n+2}\starcom{A_\mu}{A_\nu^{(n)}} - \frac{i}{n+2}\starcom{A_\nu}{A_\mu^{(n)}}
                       + \frac{i}{n+2}\starcom{\alpha}{\partial_\mu A_\nu^{(n)} - \partial_\nu A_\mu^{(n)}}\\
                   &\quad&  - \sum_{k=0}^n i \starcom{A_\mu^{(k)}}{A_\nu^{(n-k)}}\\
\end{eqnarray*}
Using the definition of $F_{\mu\nu}^{(n)}$
$$ \partial_\mu A_\nu^{(n)} - \partial_\nu A_\mu^{(n)}
   = F_{\mu\nu}^{(n)} + \sum_{k=0}^{n-1} i
   \starcom{A_\mu^{(k)}}{A_\nu^{(n-1-k)}} \; ,
$$
the Jacobi identity
\begin{eqnarray*}
  \starcom{\alpha}{i \starcom{A_\mu^{(k)}}{A_\nu^{(n-1-k)}}}
    &=& \starcom{A_\mu^{(k)}}{i \starcom{\alpha}{A_\nu^{(n-1-k)}}}
      - \starcom{A_\nu^{(n-1-k)}}{i \starcom{\alpha}{A_\mu^{(k)}}}\\
    &=& (n-k+1) \starcom{A_\mu^{(k)}}{A_\nu^{(n-k)}}
      + (k+2) \starcom{A_\mu^{(k+1)}}{A_\nu^{(n-1-k)}} \; ,
\end{eqnarray*}
and replacing in $F_{\mu\nu}^{(n+1)}$, most of the terms cancel, leaving
\begin{equation}
F_{\mu\nu}^{(n+1)} = \frac{i}{n+2} \starcom{\alpha}{F_{\mu\nu}^{(n)}} = 0
\end{equation}

%%%%%%%%%%%%%%%%%%%%%%%%%%%%%%%%%%%%%%%%%%%%%%%%%%%%%%%%%%%%%%%%%%%%%%%%%%%%%%%
%                              NON-ABELIAN THEORY                             %
%%%%%%%%%%%%%%%%%%%%%%%%%%%%%%%%%%%%%%%%%%%%%%%%%%%%%%%%%%%%%%%%%%%%%%%%%%%%%%%
\setcounter{equation}{0}
\section{Non-abelian theory}
\subsection{The pure gauge} \label{sbs:NApg}
To go on with the pure gauge in the non-abelian theory, we have to find some parametrisation compatible with the gauge transformation
$$ \delta A_\mu = \partial_\mu \lambda + i [\lambda ; A_\mu] $$
Indeed, the crucial point in the abelian theory was the existence of $\alpha$, such that
$A_\mu = \partial_\mu \alpha$ and $\lambda = \delta \alpha$. And we already know a generalization of this, allowing the non-abelian gauge transformation:
\begin{equation} \label{eq:NAbCPG}
\begin{array}{ccccccccl} \vspace{2mm}
 A_\mu   &=& \partial_\mu \alpha &+& {\displaystyle \frac{i}{2}\: [\alpha;\partial_\mu \alpha]}
         &+& {\displaystyle \frac{i}{6}\: [\alpha;i[\alpha;\partial_\mu \alpha]]} &+& \dots        \\
 \lambda &=& \delta \alpha       &+& {\displaystyle \frac{i}{2}\: [\alpha;\delta \alpha]}
         &+& {\displaystyle \frac{i}{6}\: [\alpha ; i[\alpha;\delta \alpha]]}     &+& \dots
\end{array} \end{equation}
These are exactly the formulas (\ref{eq:AbRecPG}) for $\hat{A}_\mu$ and
$\hat{\lambda}$, the commutator being now a gauge commutator instead
of a star commutator. Applying the results, without the hats and the
stars, we see that the gauge transformation of a non-abelian $A_\mu$
is the expected one, and that the field strength vanishes. We can also
point out that all gauge commutators vanish in the abelian case, and
that $A_\mu$ and $\lambda$ then reduce to the shorter expected
formulas.\\

On the noncommutative side, we still have the expressions
\begin{equation} \begin{array}{ccccccccl} \vspace{2mm}
 \hat{A}_\mu   &=& \partial_\mu \alpha &+& {\displaystyle \frac{i}{2}\: \starcom{\alpha}{\partial_\mu \alpha}}
               &+& {\displaystyle \frac{i}{6}\: \starcom{\alpha}{i \starcom{\alpha}{\partial_\mu \alpha}}} &+& \dots \\
 \hat{\lambda} &=& \delta \alpha       &+& {\displaystyle \frac{i}{2}\: \starcom{\alpha}{\delta \alpha}}
               &+& {\displaystyle \frac{i}{6}\: \starcom{\alpha}{i \starcom{\alpha}{\delta \alpha}}}       &+& \dots
\end{array} \end{equation}
where the commutators take care at the same time of the
noncommutativity and of the non-abelian gauge group. The relation between
the above $A_\mu$ and $\hat{A}_\mu$ is precisely the SW map.\\
In particular, we get $A_\mu$ and $\lambda$ for $\theta=0$, as
expected. So that the expressions of the noncommutative variables in terms
of the commutative ones, order by order in $\theta$, start like
\begin{eqnarray*}
 \hat{A}_\mu   &=& A_\mu + A_\mu^{(1)} + \dots \\
 \hat{\lambda} &=& \lambda + \lambda^{(1)} + \dots
\end{eqnarray*}

To understand why such a parametrisation exists, and what it means, let us see that we can go beyond this perturbative
development. If $A$ is a pure gauge, then it can be written in terms of an element $g=e^{-i\alpha}$ of the gauge group
$$ A = i g^{-1} dg $$
Developed in powers of $\alpha$, this expression is nothing but (\ref{eq:NAbCPG}). The non-perbutative formula for $\lambda$ is $\lambda = i g^{-1} \delta g$ (which is nothing but the gauge transformation of $g$).
Of course, it can be checked directly that $F$ is zero or that the right gauge transformation holds.\\
On the noncommutative side, the formulas are the same. We use the corresponding element
$\displaystyle{\hat{g} = (\star e)^{-i\alpha} = 1 - i\alpha - \frac{\alpha\star\alpha}{2} + \dots}$,
and replace ordinary multiplications by star products
\begin{eqnarray}
 \hat{A}       &=& i \hat{g}^{-1} \star d\hat{g}       = d\alpha + \frac{i}{2} \starcom{\alpha}{d\alpha}
                               + \frac{i}{6} \starcom{\alpha}{i\starcom{\alpha}{d\alpha}} + \dots \nonumber\\
 \hat{\lambda} &=& i \hat{g}^{-1} \star \delta \hat{g} = \delta \alpha + \frac{i}{2} \starcom{\alpha}{\delta \alpha}
                               + \frac{i}{6} \starcom{\alpha}{i\starcom{\alpha}{\delta \alpha}} + \dots
\end{eqnarray}

\subsection{The general case} \label{sbs:NAG}
Is it possible to generalize the recursive picture from abelian theory
to non-abelian one~? To do this, we first have to find a ``nice"
solution at order $\theta^2$, which means that it should at least
reduce to the abelian solution that we found in \ref{sbs:ab2o}. But with
the appearance of gauge commutators, there is a lot of freedom in
generalizing, and we face some new problems:
\begin{itemize}
\item[--] one term out of two in fig.\ref{fig:Diag} was a vanishing
  commutator, now it does not vanish,
\item[--] the derivative could be replaced by the covariant
  derivative or stay as an ordinary derivative,
\item[--] the order in $A$ (the lines) is not well defined, now that
  the field strength, the gauge transformation and the covariant
  derivative have a quadratic part.
\end{itemize}
The last point is the reason why we have to calculate the columns
first (see fig.\ref{fig:Diag}), and sort the terms thereafter,
identifying which term belongs to which line.\\

Let us start with order $\theta$. The Seiberg-Witten equation is
$$
\delta A_\mu^{(\infty,1)} - i [\lambda;A_\mu^{(\infty,1)}]
  - \partial_\mu \lambda^{(\infty,1)} + i [A_\mu;\lambda^{(\infty,1)}]
             = - \frac{1}{2} \{ \partial\lambda \stackrel{\theta}{,} \partial A_\mu \}
$$
We can point out that here $\delta$ is non-abelian, but its
non-abelian part is partly removed by the second term. The two
other terms build up the covariant derivative of
$\lambda^{(\infty,1)}$. It has the solution
\begin{eqnarray}
 \lambda^{(\infty,1)} &=& -\frac{1}{4} \{ A \stackrel{\theta}{,} \partial \lambda \} \nonumber\\
 A_\mu^{(\infty,1)}   &=& -\frac{1}{4} \{ A \stackrel{\theta}{,} \partial A_\mu + F_{\cdot\mu} \} \nonumber\\
 F_{\mu\nu}^{(\infty,1)} &=& -\frac{1}{2} \{ F_{\mu\cdot} \stackrel{\theta}{,} F_{\cdot\nu} \}
                      -\frac{1}{2} \{ A \stackrel{\theta}{,} \frac{\partial + D}{2} F_{\mu\nu} \}
\end{eqnarray}
where $D$ is the gauge covariant derivative: $D = \partial - i[A;\cdot]$\\
As already emphasized, in the abelian case all these terms were of
order $A$ (meaning $A\lambda$ or $AA_\mu$). Here they go to order
$A^3$. The question is whether they all belong to the first ``line"
anyway, higher orders in $A$ being hidden in $D$ and $F$, or not.
\\

At order $\theta^2$, the equation is
$$ \begin{array}{l} \displaystyle
\delta A_\mu^{(\infty,2)} - i [\lambda;A_\mu^{(\infty,2)}]
  - \partial_\mu \lambda^{(\infty,2)} + i [A_\mu;\lambda^{(\infty,2)}]
     =\\[2mm] \qquad\qquad \displaystyle
         i [\lambda^{(\infty,1)};A_\mu^{(\infty,1)}]
         - \frac{1}{2} \{ \partial\lambda^{(\infty,1)} \stackrel{\theta}{,} \partial A_\mu \}
         - \frac{1}{2} \{ \partial\lambda \stackrel{\theta}{,} \partial A_\mu^{(\infty,1)} \}
         - \frac{i}{8} [ \partial\partial\lambda \stackrel{\theta\theta}{,} \partial\partial A_\mu ]
\end{array} $$
for which solutions are known \cite{JMSSW,GH,BCPVZ}. We have found
another one, getting closer to the more or less expected form, but
only fitting it partially. If we try to guess a recursive formula, we
again face the problem of identifying order (1,1) and (2,2) among
their respective columns. For example, the conjectured generalization
for $F_{\mu\nu}^{(2,2)}$ could be
\begin{equation}
F_{\mu\nu}^{(2,2)} \:\stackrel{??}{=}\: -\frac{1}{2} \{ F_{\mu\cdot}^{(1,1)} \stackrel{\theta}{,} F_{\cdot\nu} \}
                            -\frac{1}{2} \{ A^{(1,1)} \stackrel{\theta}{,} \frac{\partial + 2 D}{3} F_{\mu\nu} \}
\end{equation}
Paradoxically enough, the first part, which was harder to check in the abelian case, gives exactly the right result, while there are still problems with the second part.\\

%%%%%%%%%%%%%%%%%%%%%%%%%%%%%%%%%%%%%%%%%%%%%%%%%%%%%%%%%%%%%%%%%%%%%%%%%%%%%%%
%                       ACKNOWLEDGEMENTS AND BIBLIOGRAPHY                     %
%%%%%%%%%%%%%%%%%%%%%%%%%%%%%%%%%%%%%%%%%%%%%%%%%%%%%%%%%%%%%%%%%%%%%%%%%%%%%%%
\vspace{2\baselineskip}
\begin{center} \bf ACKNOWLEDGMENTS \end{center}
I would like to thank A.~Armoni and R.~Minasian for useful discussions
and advice.

\end{document}